\documentclass[journal]{IEEEtran}
\usepackage{amsfonts}
\IEEEoverridecommandlockouts

\ifCLASSINFOpdf
\else
\fi
\usepackage{epstopdf}
\usepackage{multirow}
\usepackage{cite}
\usepackage{stfloats}
\usepackage{color}
\usepackage{epsfig}
\usepackage{graphicx}  
\usepackage{float}  
\usepackage{subfigure}  
\usepackage{psfig}
\usepackage{epsf}
\usepackage{slashbox}
\usepackage{amssymb }
\usepackage[cmex10]{amsmath}
\usepackage{algorithm} 
\usepackage{algorithmic}
\usepackage{booktabs}
\usepackage{amsmath,bm}
\usepackage{colortbl}
\usepackage{caption}
\usepackage{makecell}
\begin{document}
\title{Revolutionizing Symbiotic Radio: Exploiting Tradeoffs in Hybrid Active-Passive Communications}
\author{Rui Xu, Yinghui Ye, Haijian Sun, Liqin Shi, and Guangyue Lu \vspace{-15pt}
\thanks{Rui Xu, Yinghui Ye, Liqin Shi, and Guangyue Lu are with the Shaanxi Key Laboratory of
Information Communication Network and Security, Xi'an University of Posts \& Telecommunications, China (e-mails: dhlscxr@126.com, connectyyh@126.com, liqinshi@hotmail.com, tonylugy@163.com).}
\thanks{Haijian Sun is with  the School of Electrical and Computer Engineering, The University of Georgia, Athens, GA, USA (e-mail: hsun@uga.edu). }
}
\markboth{}
{Shi\MakeLowercase{\textit{et al.}}:}
\maketitle

\begin{abstract}
Symbiotic radio (SR), a novel energy- and spectrum-sharing paradigm of backscatter communications (BC), has been deemed a promising solution for ambient Internet of Things (A-IoT), enabling ultra-low power consumption and massive connectivity.
However, A-IoT nodes utilizing BC suffer from low transmission rates, which may limit the applications of SR in A-IoT scenarios with data transmission requirements.
To address this issue, in this article, we introduce
hybrid active-passive communications (HAPC) into SR by exploiting tradeoffs between transmission rate and power consumption.
We first present an overview of novel BC paradigms including ambient BC and SR. Then, a novel HAPC-enabled SR is proposed to enhance the transmission rate of A-IoT nodes. Furthermore,
within this paradigm, we investigate the resource allocation scheme and present preliminary research results. Simulation results show
that the transmission rate of A-IoT nodes in the proposed  HAPC-enabled SR surpasses that in traditional SR. Finally, we discuss open issues related
to HAPC-enabled SR.
\end{abstract}
\vspace{-5pt}
\IEEEpeerreviewmaketitle
\section{Introduction}

\IEEEPARstart{D}{riven} by the diverse applications of the next generation network, such as advanced transportation and smart healthcare, the rapidly growing lightweight Internet-of-Things (IoT) devices exhibit demands for
ultra-low power consumption and massive connectivity.
To address these challenges, 3GPP has shifted its focus to ambient IoT (A-IoT) \cite{1234},  and a new
RAN-level study item on A-IoT was approved in the
3GPP TSG-RAN \#97 meeting.
 A-IoT promises to significantly reduce power consumption compared to existing low-power wide-area technologies, while supporting a higher device density than that of the current IoT solutions.
This motivates us to call for high energy efficiency (EE) and spectrum efficiency (SE) communication paradigms.

In recent years, ambient backscatter communications (AmBC) \cite{8368232} has attracted much attention, in which the A-IoT device, equipped with the backscatter circuit, transmits information passively by riding on the received legacy signal emitted by existing ambient radio
frequency (RF) sources, such as cellular base stations or TV towers, and harvests energy from the legacy signal to sustain its circuit operations simultaneously.
Without generating carrier wave signals, the A-IoT device consumes significantly less power compared to the device with traditional active communications (AC), thereby achieving a high level of EE.
However, due to the non-cooperative spectrum-sharing mechanism between the legacy and backscatter transmissions, there will be co-channel interference between them, which may deteriorate transmission performance and pose significant challenges to its applications in A-IoT.
To address this issue, a novel symbiotic radio (SR) \cite{8907447} that exploits a cooperative spectrum-sharing mechanism into AmBC has emerged.
To be specific, thanks to the synchronous between the legacy and backscatter transceivers achieved by the cooperative spectrum-sharing mechanism,  the legacy receiver is capable of  transforming  backscattered signals, which contain the legacy information but are treated as  an interference in AmBC,  into beneficial multipath components  for the legacy transmission, and in return, the A-IoT receiver can employ successive
interference cancellation (SIC) to remove the legacy signal before decoding it's signals, thus achieving highly reliable
BC without requiring additional spectrum and power-consuming active components.
Accordingly, both transmissions in SR establish a mutually beneficial relationship that facilitates the effective sharing of spectrum and energy resources, thereby enhancing both SE and EE.


In this article, we first provide an overview of novel BC paradigms, including AmBC and SR, and then point out the limitations of SR for its applications in A-IoT.
To address the existing issues in SR, we first introduce the rationale behind hybrid active-passive communications (HAPC), highlight the potential of integrating HAPC with SR, and then propose a novel BC paradigm, namely HAPC-enabled SR.
Following that, we delve into the resource allocation scheme of the proposed HAPC-enabled SR and present simulation results to demonstrate its superiority.
 Finally, some open research issues are discussed.

%
%
%
%
%
%
%
%
%
%

\section{Novel BC Paradigms}

Owing to its ultra-low power consumption, BC has been recognized as a crucial technology to alleviate energy shortage in IoT. However, the conventional BC paradigms, i.e., monostatic BC and bistatic BC, heavily rely on additional spectrum resources, posting challenges for massive connections demanded by A-IoT.
Amidst the ongoing technological revolution, novel BC paradigms, such as AmBC and SR, have garnered considerable research interest, emerging as promising cornerstone technologies for A-IoT.
In this section, we focus on AmBC and SR, providing a comprehensive introduction and development analysis of these paradigms.

\subsection{AmBC}
AmBC is an energy- and spectrum-sharing paradigm, in which the A-IoT device,
equipped with the backscatter circuit, communicates passively by leveraging the incident legacy signal emitted by ambient RF sources, such as cellular base stations, TV towers, and Wi-Fi access points.
The basic principle of AmBC lies in that the A-IoT device tunes its antenna load impedance to adjust the backscatter coefficient, thereby controlling the amplitude and phase of the backscattered signal. As depicted in Fig. 1 (a), the incident legacy signal is partitioned into two parts by the backscatter coefficient:
one serves as the carrier signal, upon which the A-IoT device modulates its information and reflects the modulated signal to its associated receiver, while the remaining portion is fed into the energy harvesting (EH) circuit to sustain its operations.
Unlike traditional monostatic and bistatic BC, AmBC does not necessitate dedicated RF signal generators and extra spectrum resources, reducing deployment costs and enhancing SE.

So far, AmBC has garnered considerable attention from both industry and academia, prompting extensive research efforts, mainly including: 1) prototyping on diverse ambient signals \cite{talla2021advances}, such as Wi-Fi signals, LoRa, BLE, or ZigBee, to develop backscatter systems with varied features;
2) algorithm development in channel estimation \cite{9222226}, signal detection \cite{8007328}, etc.
However, due to non-cooperative spectrum-sharing mechanism,
uncontrollable legacy signals impose strong interference on the weak backscattered signal experiencing double path-loss fading, thereby limiting the transmission rate of the A-IoT device in AmBC. Mutually, the backscattered signal also brings interference to the legacy signal, which may deteriorate the transmission performance of ambient sources and then reduce its enthusiasm
to participate in collaboration.
Hence, mitigating the existing co-channel interference in AmBC is crucial for promoting its application in A-IoT.

\subsection{SR}

To address the limitations of AmBC, SR, which relies on the cooperation between both the legacy and backscatter transmissions, has been proposed.
Specifically,
the communication paradigm for a basic SR is depicted in Fig. 1 (b), which consists of two transmission processes: the legacy transmission and the backscatter transmission.
In the legacy transmission, the ambient source transmits its signals to the receiver through AC, while in the backscatter transmission, the A-IoT device switches
its load impedance to change the amplitude and/or phase of
its backscattered signal, and thereby modulates its information
bits on the received legacy signal. At the receiver, both the legacy and backscattered signals are received simultaneously, where the backscattered signal contains the information of both the ambient source and A-IoT device and, in general, the modulation rate of the A-IoT device is much slower than that of the ambient source as shown in Fig. 1 (b). It is worth noting that such characteristics of the received signal are also present in AmBC, while the legacy and backscatter receivers cannot fully take advantage of these characteristics due to its non-cooperative spectrum-sharing mechanism.

Unlike AmBC, benefiting from the cooperative spectrum-sharing mechanism in SR, the synchronous between the legacy and backscatter transmissions can be realized, and the channel state information of both the legacy and backscatter links is known for the receiver to support joint decoding.
Accordingly, the receiver boasts the ability to transform the backscatter signal, which contains the ambient source's information but is treated as interference in
AmBC, into beneficial multipath components for the legacy transmission.
Based on this, the joint decoding is exploited at the receiver to decode the
information of the ambient source and A-IoT device. Specifically, the receiver first decodes the ambient source's information by regarding the backscattered signal as a beneficial multipath rather than interference, and then removes the legacy signal from the received signal to decode the A-IoT device's information by using SIC.
By doing so, the co-channel interference issue existing in AmBC can be effectively solved.
Overall, the A-IoT device provides multipath gain, rather than the interference, to the legacy transmission, and in return, it shares the spectrum and energy resources of the legacy transmission achieving a reliable BC, thereby
forming a mutualistic relationship
between the legacy and backscatter transmissions in SR.

The mutual benefit nature has sparked research interests in SR, mainly from the following four aspects: 1) mutualistic relationship verification in \cite{8907447} and mutualistic condition exploration in \cite{9751388};
2) rate analysis in \cite{8399824} and capacity analysis in \cite{10214560};
3) resource allocation schemes regarding the weighted sum rate maximization of backscatter devices in \cite{8665892}, system EE maximization in \cite{9461158}, and so on, by jointly optimizing various communication resources.
4) combination with reconfigurable intelligent surface in \cite{9652042}.
The above works show that SR can achieve mutualism between the legacy and
backscatter transmissions for higher EE and SE.
Compared with AmBC, SR converts the interference caused by the A-IoT device into a beneficial multipath component for the legacy transmission and suppresses the strong interference caused by the legacy signal through SIC technique.
\begin{figure} 
	\centering  
	\vspace{-0.35cm} 
	\subfigtopskip=2pt 
	\subfigbottomskip=2pt 
	\subfigcapskip=-5pt 
	\subfigure[AmBC]{
		\label{subfig1}
		\includegraphics[width=0.95\linewidth]{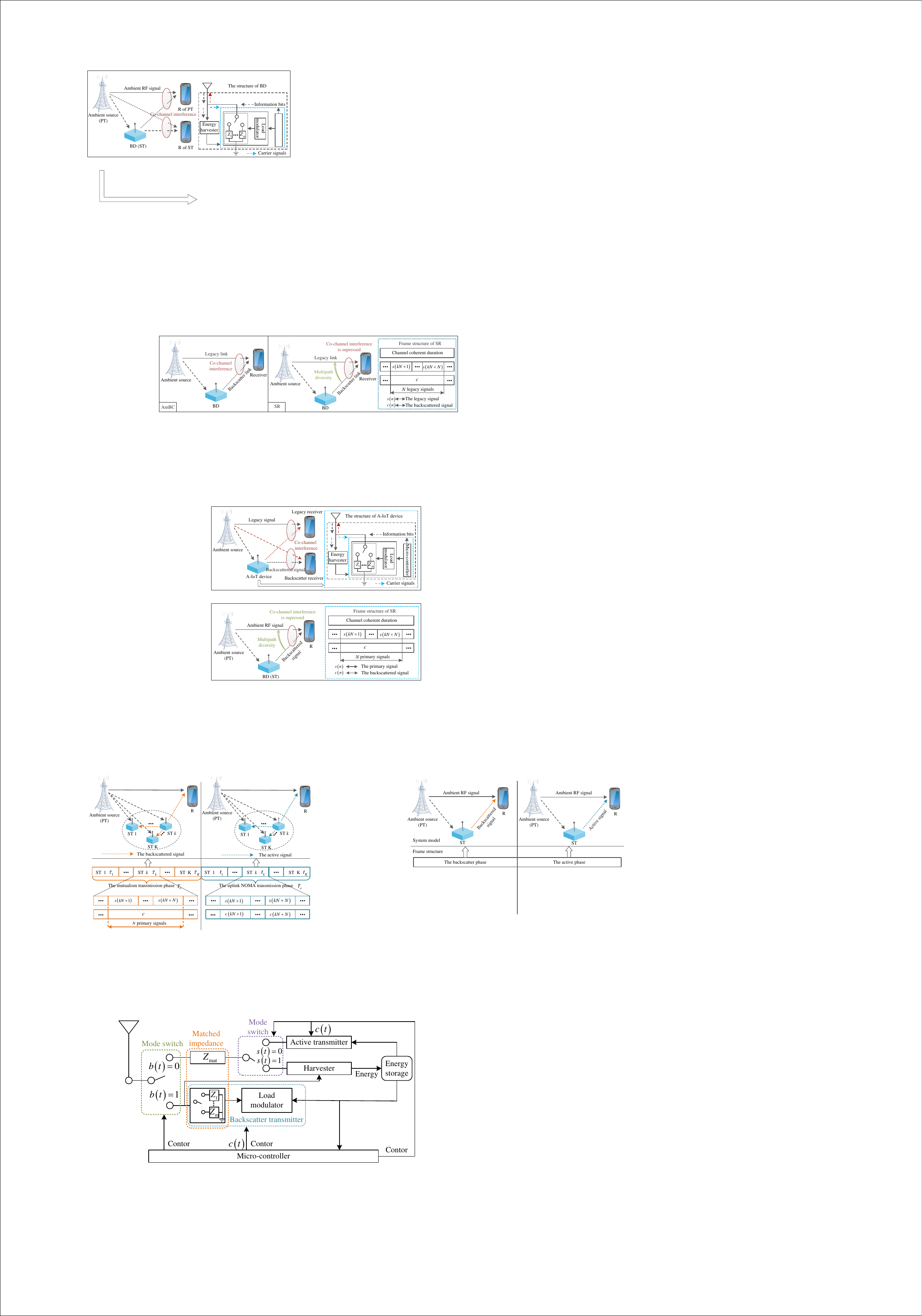}}

	\subfigure[SR]{
		\label{subfig2}
		\includegraphics[width=0.95\linewidth]{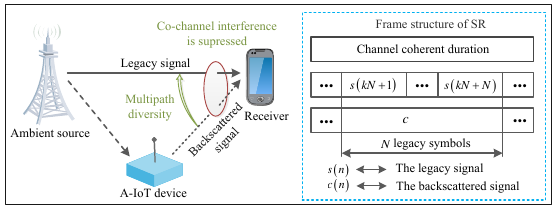}}
	
	\caption{\small{Novel paradigms for BC.}}
	\label{fig1}
\end{figure}

Although SR boasts the above advantages, to promote its application in A-IoT, there are still limitations needed to be addressed.
 Particularly, the achievable transmission rate of the A-IoT device in SR is limited due to the discrepancy in modulation rates between the A-IoT device and the ambient source, as well as the double path-loss fading experienced by the backscattered signal.
       The limited transmission rate of A-IoT devices in SR may not satisfy the rate demands of A-IoT at the required transmission distances, as listed in Table I \cite{3434}, and thus needs improvement.

\begin{table}[t]
\centering
\caption{{A-IoT Network Specifications}}
\label{table I}
 \renewcommand{\arraystretch}{1.2}
 \captionsetup[table]{singlelinecheck=off}
 \normalsize
\begin{tabular}{|c|c|}
    \hline
Parameter&Value\\\hline
User-Experienced Data Rate&0.1 kbps to 5 kbps\\\hline
Maximum Indoor Range&10 to 50 meters\\\hline
Maximum Outdoor Range&50 to 500 meters\\\hline
\end{tabular}
\end{table}
\section{HAPC-enabled SR Architecture}
In this section, we propose a novel paradigm of BC, called HAPC-enabled SR. In what follows, we first introduce the architecture of HAPC, then analyze the potential of integrating HAPC with SR, and elaborate on HAPC-enabled SR.

\subsection{HAPC Architecture}
As shown in Fig. 2, the architecture of the hybrid transmitter in HAPC mainly includes three components, i.e., active transmitter, backscatter transmitter, and energy harvester, which can be dynamically switched by adjusting the switches $a\left( t \right)$, $b\left( t \right)$ and matched impedance. In the following, we elaborate three operation modes of the hybrid transmitter.
\begin{itemize}
 \item EH: When $a\left( t \right) = 0$ and $b\left( t \right) = 1$, the hybrid transmitter harvests energy from the received RF signal completely.
  \item AC: When $a\left( t \right) = 0$ and $b\left( t \right) = 0$, the hybrid transmitter operates in the AC mode. Particularly, the antenna impedance is perfectly matched with the load impedance, and the micro-controller feeds the information $c\left( t \right)$ to the active transmitter.
  \item BC: When $a\left( t \right) = 1$, the hybrid transmitter generates different backscatter coefficients by choosing different load impedances.
 Then, the received RF signal is divided into two parts by the backscatter coefficient: one part is used as the carrier signal, upon which the hybrid transmitter  performs low-power passive BC, while
the remaining portion is fed into energy harvester.
\end{itemize}

The appeal of HAPC resides in its ability to dynamically switch among the above three modes, fully exploiting the complementary nature of AC and BC in terms of the transmission rate and power consumption, to meet
the distinct demands of various IoT scenarios.
In specific, if the transmitter has no information to transmit, it collects energy from the received RF signal. Otherwise, the device transmits information via HAPC, where it can not only perform low-power BC, but also convert to AC to improve its transmission rate when enough energy is harvested.



\begin{figure}
  \centering
  \includegraphics[width=0.45\textwidth]{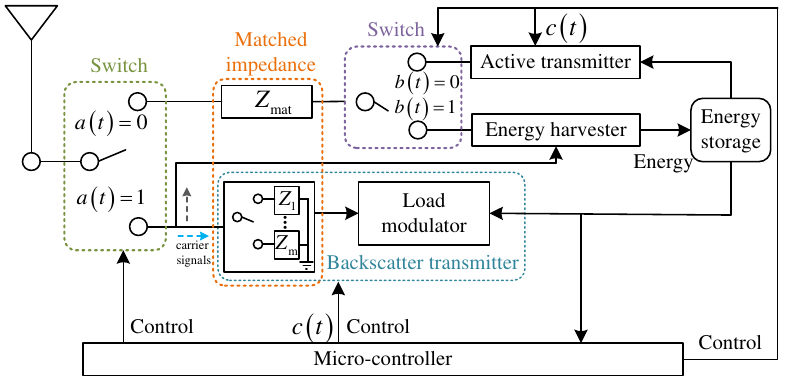}\\
  \caption{\small{The structure of the hybrid transmitter in HAPC.}}
  \label{fig2}
\end{figure}
\subsection{The Potential of Integrating HAPC with SR}
Here, we summarize the strengths and weaknesses of HAPC and SR, and then highlight the motivation to integrate them as a whole.
\begin{itemize}
  \item Compared to AmBC, SR can eliminate co-channel interference through the cooperation mechanism between the legacy and backscatter transmissions, forming a mutually beneficial transmission while sharing spectrum and energy resources between both the A-IoT device and ambient source. However, since the A-IoT device in SR only performs BC, its modulation rate is significantly lower than AC and may not effectively support various IoT applications with certain data transmission requirements.
  \item Compared to passive BC, HAPC can effectively enhance the transmission rate of devices by utilizing the complementary characteristics between AC and passive BC.
      However, additional spectrum resources are necessary,
      which fails to satisfy the  requirements of massive connections in A-IoT.
\end{itemize}
\begin{center}
{\emph{Can strengths be leveraged and weaknesses avoided in both HAPC and SR to enhance the transmission rate of A-IoT devices in SR?}}
\end{center}

To answer this question, a natural idea arises to integrate HAPC and SR, yielding HAPC-enabled SR, where HAPC is applied to transmit signals of A-IoT devices to improve its transmission rate by exploiting tradeoffs between  transmission rate and power consumption, while ensuring the mutually beneficial relationship can be formed between the legacy and backscatter transmissions.
This integration paradigm exhibits immense potential particularly in massive high-density A-IoT scenarios. For example, if multiple A-IoT devices transmit information through time division multiple access,
the time of each A-IoT device for EH is much longer than its transmission time, and thus the A-IoT device may not fully consume all the harvested
energy in each transmission block if it only performs low-power passive BC.
In this situation, the A-IoT device can fully utilize the harvested energy via HAPC to improve its transmission rate.

\begin{table*}[t]
\centering
\caption{{Comparisons of different communication paradigms}}
\label{table II}
 \renewcommand{\arraystretch}{1.2}
 \captionsetup[table]{singlelinecheck=off}
 \normalsize
\begin{tabular}{|c|c|c|c|c|}
    \hline
\makecell{Communication \\ paradigms}&\makecell{Transmission rate \\ of A-IoT devices}&\makecell{Energy consumption\\ of A-IoT devices}&\makecell{Extra spectrum\\ resources}\\\hline
AmBC&low&low&no\\\hline
SR&medium-low&low&no\\\hline
HAPC&medium&medium&yes\\\hline
HAPC-enabled SR&medium&medium&no\\\hline
AC&high&high&yes\\\hline
\end{tabular}
\end{table*}

\subsection{HAPC-enabled SR}
A simple HAPC-enabled SR network architecture, as shown in Fig. 3, comprises one ambient source, $K$ A-IoT devices equipped with HAPC circuit, and one receiver. In this network, the ambient source transmits its signals via AC, while $K$ A-IoT devices harvest energy from the legacy signal emitted by the ambient source to support their information transmission via HAPC. This network configuration is well-suited for various A-IoT scenarios, e.g., a telephone (ambient source) transmits its signal to a gateway (receiver), and multiple smart home sensors (A-IoT devices) want to deliver information to the gateway by riding on the telephone's signal or AC.
There are three operation modes for A-IoT devices, i.e., passive BC, AC, and EH, and the entire transmission slot is divided into two phases: the mutualism transmission phase and the uplink non-orthogonal multiple access (NOMA) transmission phase.

(1) In the mutualism transmission phase, $K$ A-IoT devices take turns to transmit their signals via BC, and harvests energy from the received legacy and backscattered signals.
Under this setting, the receiver receives the legacy and backscattered signals simultaneously, where the latter contains the information of both the ambient source and A-IoT device.
Due to the modulation rate discrepancy between the ambient source and A-IoT device, i.e., the A-IoT device only modulates one symbol onto $N$ ($N \gg 1$) symbols of the ambient source, as shown in Fig. 3, the backscatter transmission of the A-IoT device provides an additional multipath gain for the legacy transmission.
Accordingly, SIC technique can be carried out to decode them sequentially, i.e., the receiver first decodes the ambient source's information, and then removes the legacy signal from the received signal to decode the A-IoT device's information.

(2) In the uplink NOMA transmission phase, $K$ A-IoT devices convert to AC mode and access the spectrum of the ambient source alternatively as two-user uplink NOMA. Meanwhile, A-IoT devices without information transmission harvest energy from the legacy signal. At the receiver, co-channel interference exists and joint decoding with SIC in NOMA is applied.
To achieve a higher transmission rate for A-IoT devices,
the receiver first decodes the legacy signal regarding the backscattered signal as interference, which would cause the performance degradation of the ambient source. Nevertheless, the performance degradation of the ambient source in the uplink NOMA phase can be compensated by the performance gain of the ambient source in the mutualism
transmission phase through designing effective resource allocation scheme, thereby achieving a mutualistic relationship in the whole HAPC-enabled SR network.

Accordingly, the proposed HAPC-enabled SR can effectively address the limitations encountered in SR, realizing a significant enhancement for the transmission rate of the A-IoT device.
Until now, various communication paradigms have been mentioned, including AmBC, SR, HAPC, HAPC-enabled SR, and AC. In order to clearly illustrate the superiority of the proposed HAPC-enabled SR, Table II summarizes
and compares these communication paradigms.

\begin{figure}
  \centering
  \includegraphics[width=0.5\textwidth]{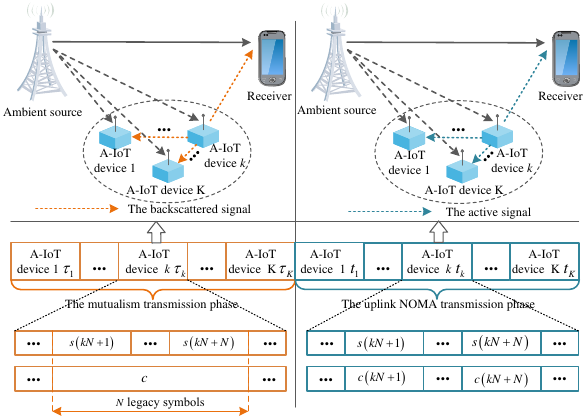}\\
  \caption{\small{The network architecture of HAPC-enabled SR}}\label{fig3}
\end{figure}

\section{Resource Allocation for HAPC-enabled SR}
\begin{figure*}
\centering
\subfigure[]{\label{fig:subfig:a}
\includegraphics[width=0.3\linewidth]{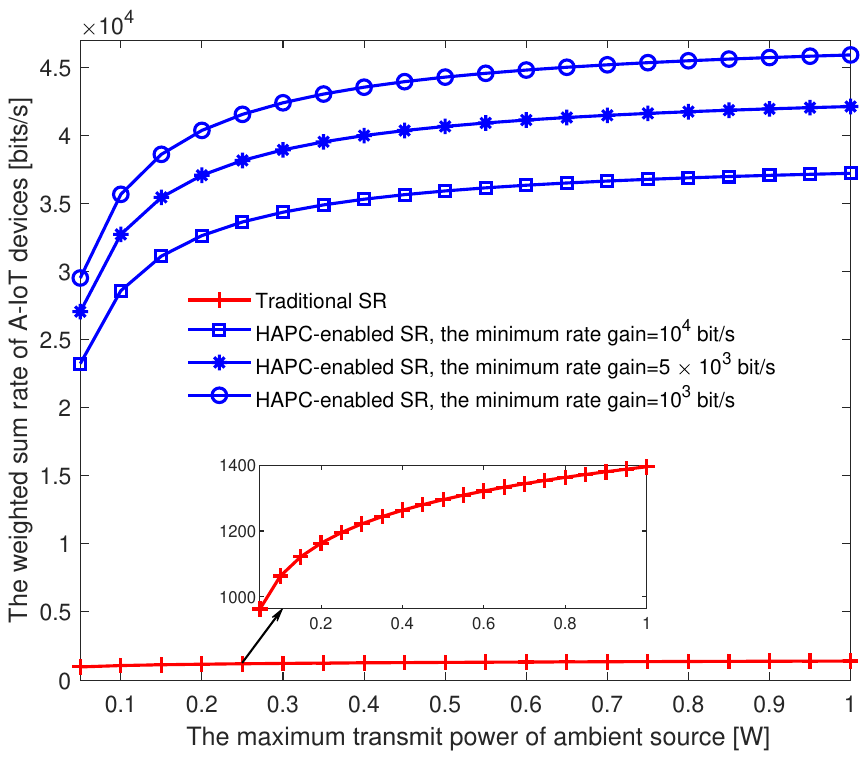}}
\hspace{0.01\linewidth}
\subfigure[]{\label{fig:subfig:b}
\includegraphics[width=0.3\linewidth]{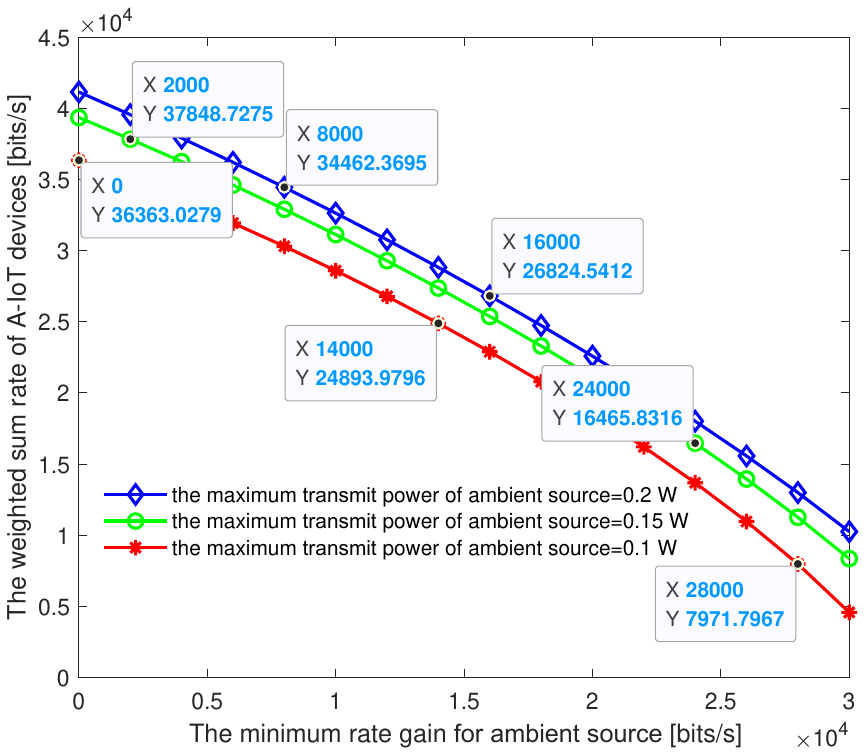}}
\subfigure[]{\label{fig:subfig:a}
\includegraphics[width=0.3\linewidth]{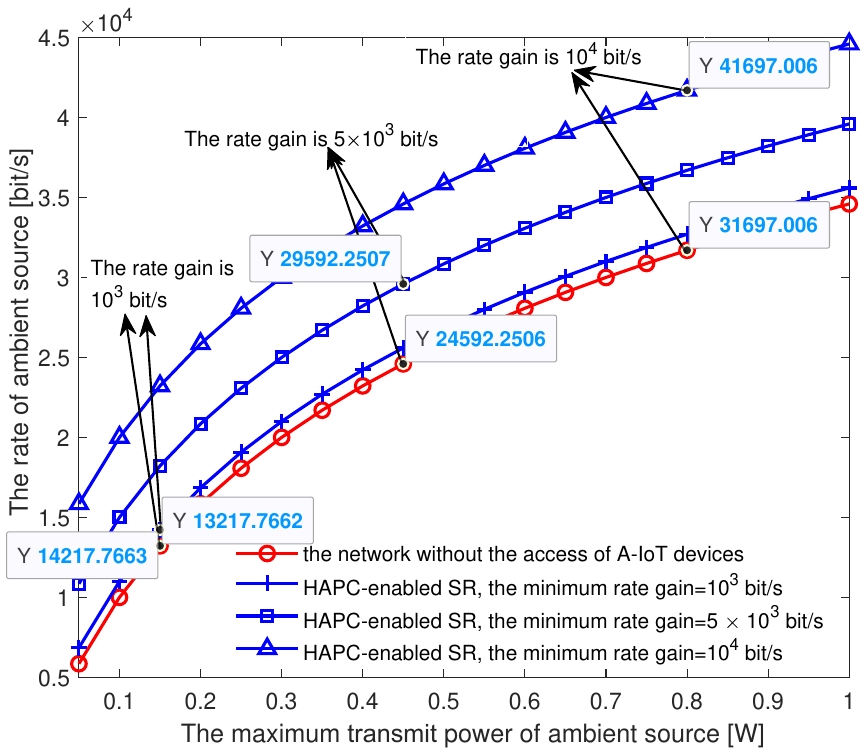}}
\hspace{0.01\linewidth}
\label{fig:subfig}
\caption{\small{Simulation results of HAPC-enabled SR. (a) The weighted sum rate of A-IoT devices versus the maximum transmit power of ambient source; (b) The weighted sum rate of A-IoT devices versus the minimum rate gain for ambient source; (c) The rate of ambient source versus its maximum transmit power.}}
\end{figure*}
In the previous section, HAPC-enabled SR is introduced, aiming to enhance the transmission rate of A-IoT devices in SR by harnessing the advantages of both SR and HAPC. More importantly, the key to such enhancement lies in understanding the tradeoffs among various communication resources in the proposed network and designing effective resource allocation strategies. To this end, in this section, we first analyze the tradeoffs in HAPC-enabled SR. To further demonstrate its superiority, we investigate a resource allocation scheme for  weighted sum rate maximization of A-IoT devices and present preliminary simulation results.
\subsection{Tradeoffs in HAPC-enabled SR}
In the proposed HAPC-enabled SR, there are two key trade-off relationships influencing the overall performance, which are summarized as follows.
\begin{itemize}
  \item The tradeoff between the rate gain of the ambient source and the rate improvement of A-IoT devices: In HAPC-enabled SR, the mutualistic relationship between the legacy and backscatter transmissions is one of its distinctive features. This means that the achievable rate of the ambient source in the proposed network should be greater than that without the access of the A-IoT device, and such rate difference of the ambient source is defined as the rate gain.
      This primarily benefits from the beneficial multipath components introduced by passive backscatter transmission.
However, to achieve a high rate gain of the ambient source, more communication resources should be allocated to the A-IoT device performing passive BC. This reduces the communication resources for the A-IoT device to perform AC, which may not effectively improve the transmission rate of the A-IoT device.
Consequently, there exists a tradeoff between the rate gain of the ambient source and the rate improvement of A-IoT devices.
  \item The tradeoff between active/passive transmission and EH of the A-IoT device: In HAPC-enabled SR, satisfying the energy-causality constraint for the A-IoT device, that is, the consumed energy of the A-IoT device to perform AC or/and BC cannot exceed its harvested energy, is a fundamental prerequisite for network sustainability.
  To improve the transmission rate of the A-IoT device, more communication resources should be allocated to perform AC, leaving less resources for passive BC. Under this situation, more energy is needed to support AC of the A-IoT device, which may not satisfy the energy-causality constraint.
Consequently, there exists another tradeoff between the information transmission and EH.
\end{itemize}
\subsection{Resource Allocation Schemes}
Given the above tradeoffs among communication resources within the proposed HAPC-enabled SR network under consideration, devising effective resource allocation schemes according to distinct requirements are pivotal. To illustrate the superior performance of the proposed HAPC-enabled SR, we have
explored the resource allocation scheme for weighted sum rate
maximization of A-IoT devices
in preliminary research, based on a basic system model as shown in Fig. 3.

Towards this end, we formulated a weighted sum rate maximization problem in terms of A-IoT devices, subject to the following constraints: 1) the rate gain constraint of the ambient source, meaning that the achievable rate gain of the ambient source should exceed its expected minimum rate
gain;
 2) the energy-causality constraint of A-IoT devices, indicating that the consumed energy of each A-IoT device should not exceed its harvested energy;
 3) the rate constraint of A-IoT devices, ensuring that the transmission rate of each A-IoT device is greater than zero, allowing each A-IoT device to share the spectrum resources of the ambient source for information transmission, along with other constraints on time and power resources.
 Note that constraints 1) and 3) guarantee the mutualistic relationship between the legacy and backscatter transmissions in the proposed network.
 By solving the formulated optimization problem, we finally  obtain an efficient resource allocation scheme. For brevity, the details for solving are omitted here, and the corresponding simulation results are presented below.

\subsection{Simulation Results}
In light of the above weighted sum rate
maximization problem, we present simulation results \cite{10437703} to validate that the proposed HAPC-enabled SR network can significantly improve the transmission rate of A-IoT devices.
To facilitate analysis, we consider the proposed network with two A-IoT devices and the locations of all devices are described with a 2-D Cartesian
coordinate system, in which the coordinates of the ambient source, A-IoT device 1, A-IoT device 2, and the receiver are set as (0,0) m, (0.8,0) m, (0,1) m and (100,1) m, respectively.  The basic parameter values are set as follows: the
channel bandwidth is 10 KHz, the noise
power spectral density is $-$90 dBm/Hz,
the energy conversion efficiency is 0.8, the circuit
power consumption of passive BC and AC are
${10^{ - 5}}$ W and ${10^{ - 3}}$ W.

Fig. 4 (a) shows the weighted sum rate of A-IoT devices versus the maximum transmit power of the ambient source
under the proposed HAPC-enabled SR network and the traditional SR network.
 It is observed that the weighted sum rate of A-IoT devices achieved in the proposed network is significantly larger than
that in the traditional SR network, demonstrating the superiority of the proposed network.
Additionally, the weighted sum rate of A-IoT devices decreases with the increase of the minimum rate gain for the ambient source, verifying the tradeoff between
the ambient source's rate gain and the A-IoT device's rate improvement.
Nevertheless, the weighted sum rate
of A-IoT devices under the proposed network remains higher than that under the
traditional SR network.

To further verify the tradeoff observed in Fig. 4 (a), Fig. 4 (b)  depicts the weighted sum rate of A-IoT devices versus the minimum rate gain for the ambient source.
The trend of the curve reaffirms the tradeoff between the ambient source's rate gain and the A-IoT device's rate improvement.
This is because the increase in ambient source's rate gain is accompanied by an increase in communication resources for A-IoT devices to perform BC, which undoubtedly leads to a decrease in communication resources for A-IoT devices to perform AC, resulting in a lower weighted sum rate of A-IoT devices.

Fig. 4 (c) depicts the ambient source's rate versus its transmit power under different minimum rate gains. For a given minimum rate gain, the rate gain of the ambient source, i.e., the difference
of the ambient source's rate under the proposed network and the
network without the A-IoT device's access, is equal to its minimum rate gain. This phenomenon can be explained as follows: to maximize the weighted sum rate of A-IoT devices, more communication resources should be allocated to A-IoT devices for AC, leaving fewer resources for BC. This may result in a lower rate gain for the ambient source, which may not satisfy its rate gain constraint. To ensure the minimum rate gain for the ambient source while maximizing the weighted sum rate of A-IoT devices, minimal communication resources that are just enough to meet the minimum rate gain of the ambient source should be allocated for A-IoT devices to perform BC. Consequently, as many resources as possible should be reserved for AC, thereby maximizing the weighted sum rate of A-IoT devices.
\section{Open Research Issues}
In this section, we discuss some open research issues related to the proposed HAPC-enabled SR.

\textbf{How to guarantee the mutualistic relationship:} Ensuring mutual benefit between the legacy and backscatter transmissions serves as a fundamental prerequisite for the proposed HAPC-enabled SR. However, unlike in traditional SR, the ambient source not only benefits from the multipath components provided by the backscatter transmission during the mutualism transmission phase, but also may experience performance degradation during the uplink
NOMA transmission phase. Consequently, it is of great significance to explore the conditions that satisfy mutually beneficial relationships, devise efficient resource allocation strategies, and investigate the access schemes of A-IoT devices when they perform AC, all aimed at ensuring the mutualistic relationship between the legacy and backscatter transmissions.

\textbf{Cross-layer design for HAPC-enabled SR:}
In general, the researches related to HAPC-enabled SR mainly
focus on physical layer. However, the mutual influence between the legacy and backscatter transmissions also affects the MAC-layer performance, especially when multiple ambient sources and A-IoT devices coexist while adopting the contention-based medium
access protocol.
Specifically, the backscatter transmission will increase the failure probability of the packet transmission of ambient sources in the MAC layer. In return, since A-IoT devices can only transmit information when the ambient source is active,
the performance of A-IoT devices is also related to the MAC-layer activity
of ambient sources.
Therefore, the cross-layer performance of both the legacy and backscatter system needs to be evaluated, and the cross-layer optimization should be explored to achieve the optimal user scheduling and resource allocation.

\textbf{Integration of HAPC-enabled SR and cellular networks:} Currently, the considered system model of the proposed HAPC-enabled SR faces difficulties in achieving large-scale network and unified scheduling. The integration of HAPC-enabled SR and cellular networks is the future development trend,
which can leverage the advantages of cellular infrastructure and licensed spectrum to achieve interference and flexible device management, thereby enhancing transmission coverage and reliability. Notably, the recently emerging fluid antenna can also be utilized in HAPC-enabled SR to achieve interference-free transmission through fluid antenna multiple access.

\section{Conclusions}

Both AmBC and SR enable A-IoT devices to share spectrum and energy
resources with ambient sources via passive BC.
However, the co-channel interference present in AmBC limits its applications in A-IoT, while this issue can be effectively addressed by the cooperative mechanism between the legacy and backscatter transmissions in SR, yielding a mutually beneficial relationship.
Due to the discrepancy in modulation rates between the ambient source and the A-IoT device, as well as the double path-loss
fading experienced by the backscattered signal, the transmission rate of the A-IoT device in SR is limited.
To address this issue, HAPC-enabled SR has been proposed in this article. Based on this, the related resource allocation scheme has been explored, and the preliminary simulation
results have been obtained. Simulation results demonstrate the superiority of the proposed HAPC-enabled SR. In the end,
future research issues that can be further addressed have been discussed.

\ifCLASSOPTIONcaptionsoff
  \newpage
\fi
\bibliographystyle{IEEEtran}
\bibliography{refa}

\end{document}